\documentclass[12pt,letterpaper]{article}
\usepackage{amsmath}
\usepackage{amsfonts}
\usepackage{amsthm}
\usepackage{bm}
\usepackage[labelfont = bf]{caption}
\usepackage{dsfont}
\usepackage{gensymb}
\usepackage{graphicx,psfrag,epsf}
\usepackage{enumerate}
\usepackage{multirow}
\usepackage{color}
\usepackage{changes}
\usepackage[affil-it]{authblk}
\usepackage{placeins}

\usepackage{comment}

\usepackage{natbib}

\usepackage{todonotes}
\usepackage{float}

\usepackage{url} 

\numberwithin{equation}{section}
\theoremstyle{plain}

\newcommand{\github}{\url{github.com/ClaudioHeinrich/RankHistBins}}

\newcommand{\wt}{\widetilde}

\newcommand{\Rpack}{\texttt{RankHistBins} }
\newcommand{\acct}{c_{\text{acc}}}

\graphicspath{{../../Figures/}{./}}

\begin{document}

\def\spacingset#1{\renewcommand{\baselinestretch}%
{#1}\small\normalsize} \spacingset{1.35}

  \title{\bf On the number of bins in a rank histogram}
  \author{Claudio Heinrich
 \thanks{
 email: \texttt{claudio.heinrich@nr.no}\\
 The author would like to thank Thordis Thorarinsdottir for helpful discussions, and two anonymous reviewers for their suggestions that helped to substantially improve the paper.
He thanks his colleagues from the Norwegian Computing Center for labeling many histograms and is grateful to the Norwegian Computing Center for its financial support.}}
\affil{ \footnotesize Norwegian Computing Center Oslo, \\ { P.O. Box 114 Blindern, NO-0314 Oslo, Norway}}

\maketitle

\begin{abstract}
Rank histograms are popular tools for assessing the reliability of meteorological ensemble forecast systems.
A reliable forecast system leads to a uniform rank histogram, and deviations from uniformity can indicate miscalibrations. 
However, the ability to identify such deviations by visual inspection of rank histogram plots crucially depends on the number of bins chosen for the histogram.
If too few bins are chosen, the rank histogram is likely to miss miscalibrations; if too many are chosen, even perfectly calibrated forecast systems can yield rank histograms that do not appear uniform.
In this paper we address this trade-off and propose a method for choosing the number of bins for a rank histogram. 
The goal of our method is to select a number of bins such that the intuitive decision whether a histogram is uniform or not is as close as possible to a formal statistical test. 
Our results indicate that it is often appropriate to choose fewer bins than the usual choice of ensemble size plus one, especially when the number of observations available for verification is small.
\end{abstract}

\noindent%

\section{Introduction}\label{sec:introduction}

Rank histograms are widely used diagnostic tools for calibration assessment of forecasts in meteorology. 
The underlying idea to consider the rank of the observation within a predictive ensemble was proposed independently by \citet{Anderson1996}, \citet{HamillColucci1997} and \citet{Talagrand&1997}.
If the prediction system is well-calibrated (or reliable), the rank of the observation within the ensemble is approximately uniformly distributed.
Deviations from uniformity indicate different types of miscalibration, for example, sloped histograms indicate bias, and $\cup$- or $\cap$-shaped histograms indicate under- and overdispersion, respectively.
Rank histograms were originally applied to univariate forecasts, however, several generalizations towards multivariate forecasts exist \citep{Wilks2004,Thorarinsdottir&2016,ZiegelGneiting2014}.

\begin{figure}[h]
\includegraphics[width = 0.3\textwidth]{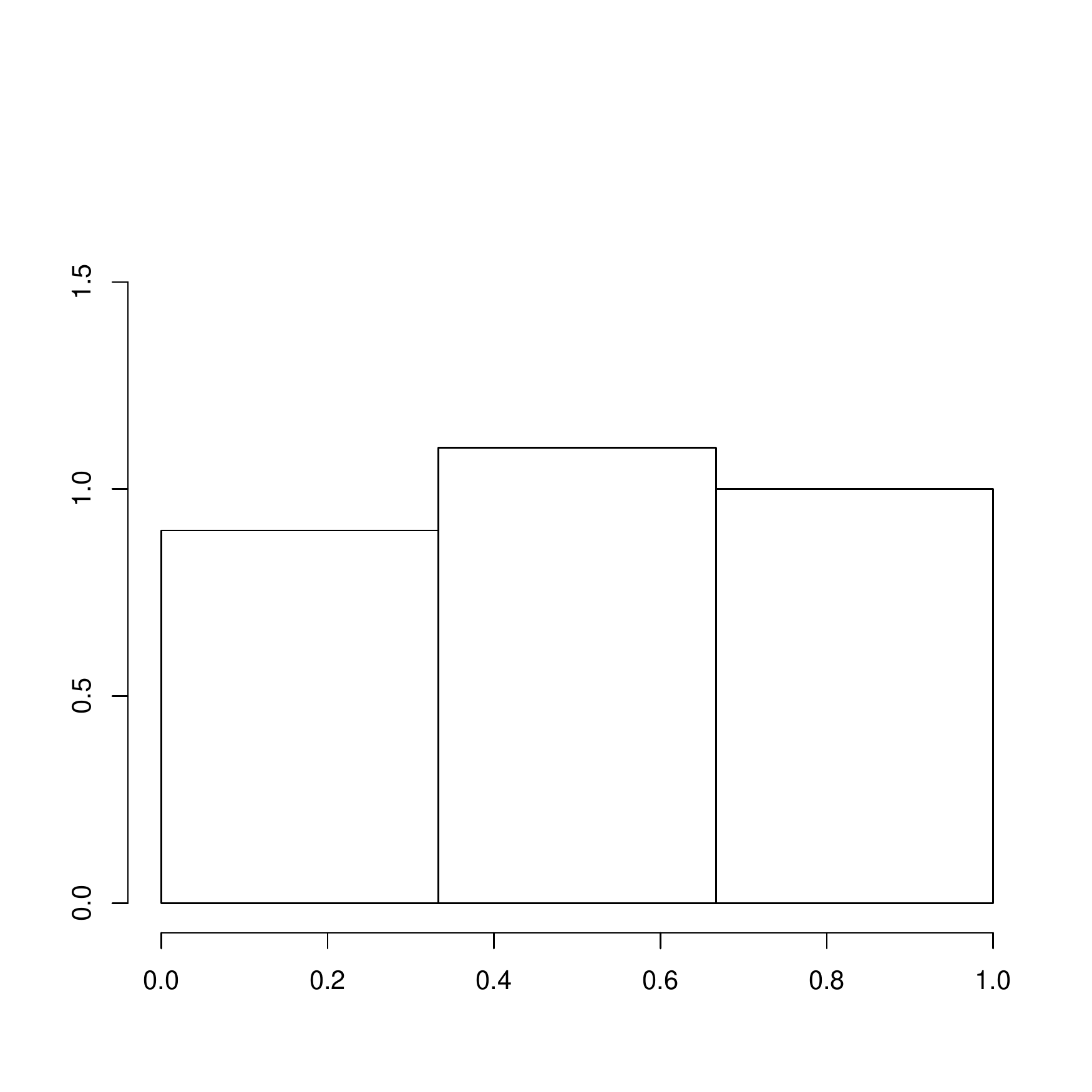}
\includegraphics[width = 0.3\textwidth]{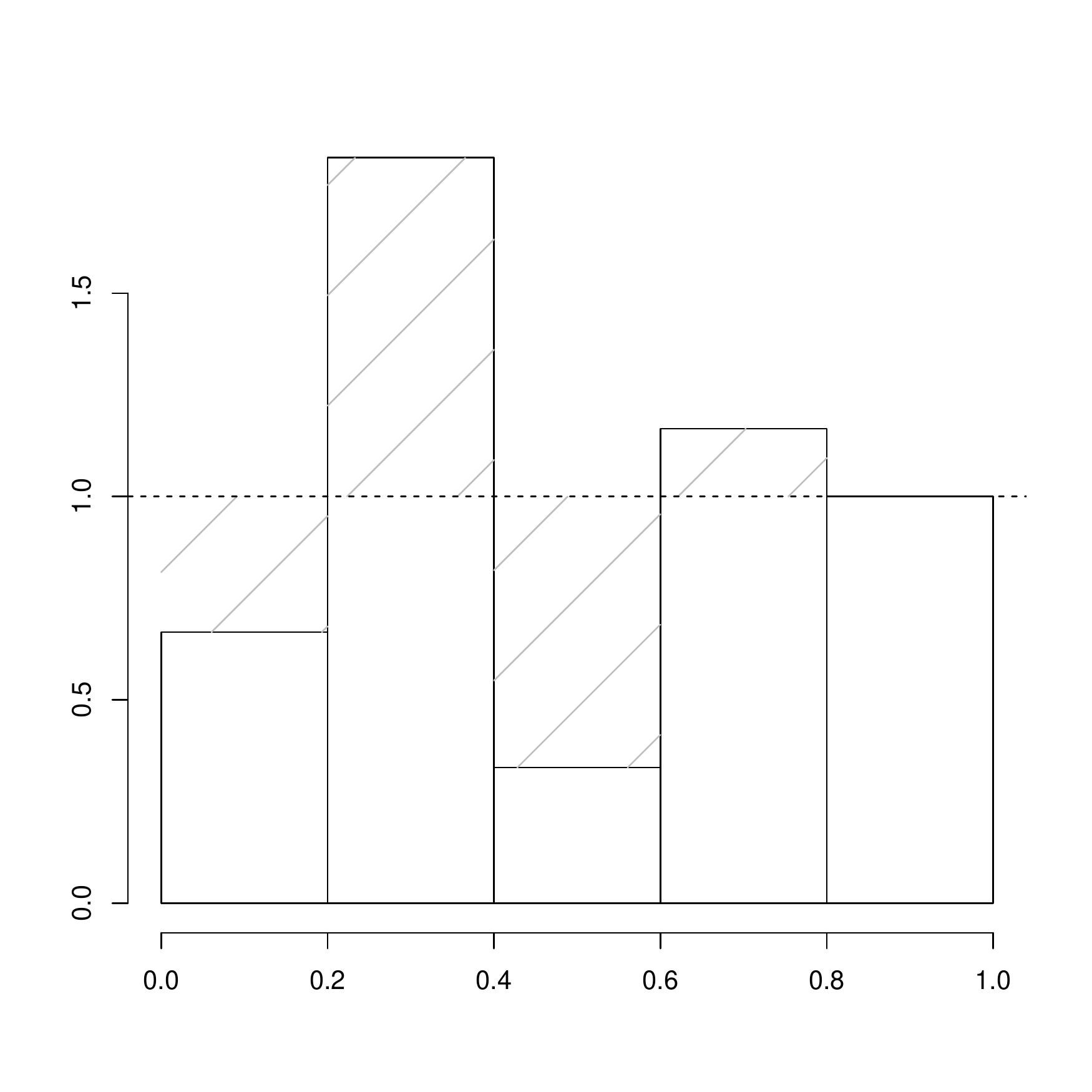}
\includegraphics[width = 0.3\textwidth]{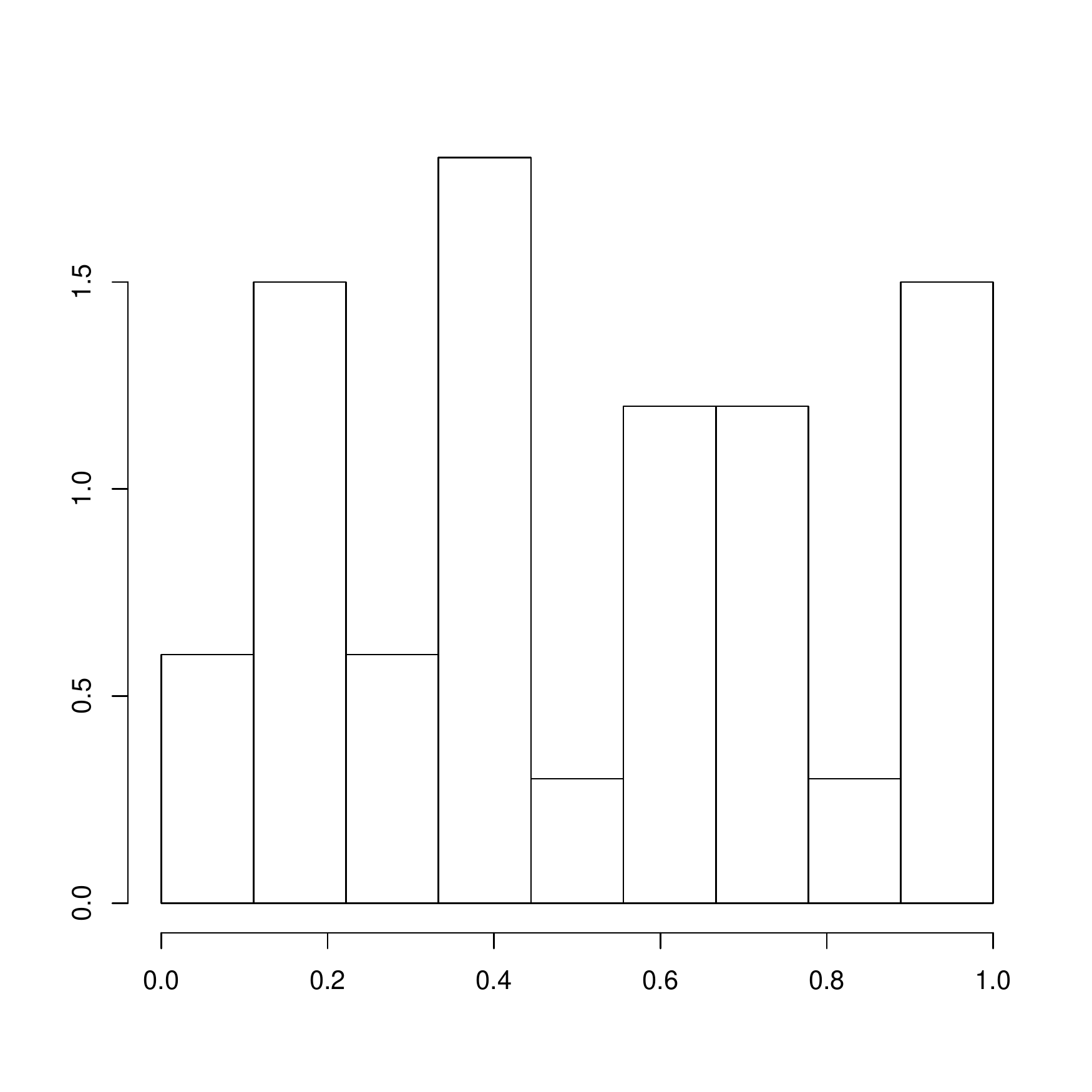}
\caption{Three histograms based on the same data with different number of bins. The data is a sample of 30 numbers, uniformly and independently distributed on [0,1]. The middle plot shows an example for a distance from uniformity considered in this paper: The size of the hatched area is the $L^1$-distance $D_{L^1}$ between the considered histogram and a perfectly flat histogram. Clearly, the distance varies with the number of bins.
\label{Fig:HistEx}}
\end{figure}

As pointed out by \citet{Wand1997} in a different context, choosing the number of bins in a histogram is generally a trade-off: More bins lead to a more detailed histogram while also making it more susceptible to random fluctuations.
In particular, when the available number of forecast-observation pairs is small, the appearance can change quite dramatically with different bin numbers, see Figure \ref{Fig:HistEx}. The goal of this work is to address this trade-off and provide guidance regarding the choice of a bin size in a rank histogram. We focus on the case where only a relatively small number of forecast-observation pairs are available, say less than 200. 
In this case, too many bins can lead to an over-interpretation of the histogram's appearance.
This situation occurs, for example, frequently in seasonal forecasting where variables are averaged over long time-spans, leading to a drastically reduced number of available observations, see \citet{VanSchaeybroeck&2018}. 

When an ensemble forecast with $m$ ensemble members is considered, the observation rank can take values between 1 and $m+1$. It is therefore intuitive and common practice to use $m+1$ bins for rank histograms, each bin corresponding to a single rank (e.g. \citet{Wilks2019}).
We show how to construct rank histograms with any bin number such that every bin accounts for the same number of ranks.
This is necessary in order to address the above-mentioned trade-off, and useful in its own right. It can, for example, be quite difficult to compare histograms with different bin numbers. Therefore, when forecast systems with different ensemble sizes are compared, it is useful to choose the same bin number for all of them.

Our approach to finding `good' bin numbers acknowledges that rank histograms are first and foremost used for exploratory data analysis. They are typically generated and inspected by scientists who then intuitively decide whether they look sufficiently uniform or not. This implies, in particular, that good bin numbers are not an inherent statistical property of the data, but require assumptions on scientists' intuitive decisions. We will assume that such decisions directly depend on the distance between the observed histogram and a perfectly flat histogram, and that larger distances are more likely to lead to a rejection.
This constitutes a necessary oversimplification, which in particular does not take characteristic shapes such as slopes or $\cup$-shapes into account.
 An empirical study is conducted where several statisticians label more than 400 histograms as uniform or not, in order to assess to what extent our assumption is justified.

Subject to this assumption, the bin number can be chosen to make the scientists' decision approximate the decision of a formal statistical test for uniformity. The underlying intuition is that, when based on uniformly distributed data, histograms with fewer bins tend to look flatter than those with many bins. Therefore, reducing the number of bins reduces the probability of an intuitive false reject (type I error). At the same time, it reduces the amount of detail depicted by the histogram and therefore increases the probability of a false accept (type II error). In this sense the trade-off in choosing the number of bins directly relates to the trade-off made in statistical testing when choosing a significance level, which balances the probabilities of the two types of errors.
We formalize this intuitive link, which then allows us to associate a chosen number of bins with a probability for a false reject. Establishing this link requires the selection of a subjective `acceptance threshold', indicating how large deviations from uniformity are deemed acceptable by the inspecting scientist. We use the results from our empirical study to provide approximations for the average scientists' acceptance threshold.

There are several different tests for uniformity that have been applied in the context of rank histograms. Besides the classical $\chi^2$-test, \cite{DelleMonache&2006} considered a test based on the so-called reliability index, and \cite{Taillardat&2016} used a test based on an entropy test statistic. These three tests have recently been compared by \citet{Wilks2019}. For all three of them, the test statistic can be interpreted as a distance between the observed and a perfectly flat histogram. This allows us to establish and analyze the above-mentioned link between the choice of bin number and a statistical test for any of the three tests.

Given a significance level $\alpha$ and the number of available observations $n$, our methodology selects 
a bin number $k$ such that, when inspecting a histogram with $k$ bins, a scientists' intuitive decision closely approximates the test at significance level $\alpha$. This bin number is in most cases similar (and often identical) for the three different tests, which provides a sanity check for our methodology: The selected bin number should lead to a false rejection by the scientist with probability $\alpha$, regardless of the test used in the derivation.

Our results show that when only few observations are available, even histograms with a moderate number of bins lead to high probabilities of an intuitive false reject.
For example, when 100 observations are available, choosing more than 9 bins results in a probability of more than $33\%$ of a false reject; for 60 available observations, this probability is exceeded when more than 6 bins are chosen. 

Optimality criteria for histogram bin numbers and bin widths have been widely discussed in the literature, see e.g. \citet{Scott1979,HeMeeden1997,Muto2019} and \cite{Knuth2019}.
However, these criteria have generally been developed in a different context and under assumptions that make them inappropriate for rank histograms.
They mostly focus on histograms as tools for estimating probability densities with the aim of finding the number of bins that
 minimizes a distance (often the mean integrated squared error) between the underlying density and the histogram of the data. 
In this context it is commonly assumed that the density is continuous and sufficiently smooth over an interval. Some early work even assumes  approximately normally distributed data \citep{Scott1979,Sturges1926}. These assumptions are not met for rank histograms based on discrete data.
Moreover, the vast majority of results derived in this strand of literature are of asymptotic nature and therefore assume $n$ to be large, in contrast to our assumptions.
Thirdly, the derived binnings are often data driven, i.e. the bin number depends on properties of the data beyond the sample size $n$, such as for example the sample variance. In the context of rank histograms, which are commonly used to compare different forecast systems this is not desirable as all the histograms should have the same number of bins.

The remainder of the paper is organized as follows. In Section \ref{sec:randomization} we show how histograms with any bin number can be derived from an $m$-member ensemble forecast. Section \ref{sec:Theory} describes the approach we take to relate the bin number to statistical tests. The optimal bin number requires the choice of a subjective acceptance threshold. In Section \ref{sec:c} we present an empirical study and use it to derive an approximation of this acceptance threshold. In Section \ref{sec:Results} we use the developed algorithm to find good bin numbers for a range of different data sizes. Section \ref{sec:non-uniform} analyzes the rejection probability for histograms with the optimal bin number under non-uniform distributions.
 Section \ref{sec:Discussion} provides a discussion of the results and Section \ref{sec:Conclusion} concludes.

\section{Changing the bin number for rank histograms}\label{sec:randomization}

When computing rank histograms for an ensemble forecast with $m$ members the observation ranks $r_1,...,r_n$ take values in $\{1,...,m+1\}$. 
Therefore, the default is to use a histogram with $m+1$ bins, each bin containing the counts for one rank only. 
It is straightforward to instead generate a rank histogram with $k<m+1$ bins, as long as $k$ divides $m+1$. Then, the first bin accounts for the first $(m+1)/k$ ranks, and so on. However, this is quite restrictive, especially as $m+1$ is prime for some popular ensemble sizes such as 10, 30 and 100. As argued in the introduction, free choice of the bin number $k$ is desirable and we show in the following how this can be achieved.

The problem that arises when $k$ does not divide $m+1$ is that some bins get assigned more ranks than others. Take the simple example of $m=2$ where the observed ranks take the values $1,2,3$, and assume we want to plot a histogram with only two bins. Then, the question arises whether the counts of rank 2 should be placed in the first or the second bin. Both options lead to skewed histograms even if the ranks are perfectly uniformly distributed.
This issue can be resolved by randomization. For each count of rank 2 we simply flip a coin and place it in the first bin if the coin shows tails, and in the second bin otherwise.
When moving beyond this simple example, the randomization becomes more involved, as it needs to account for the fraction of overlap between bins and ranks: Say, for example, we have ranks $1,...,5$ and want to consider 4 bins, then the first bin should account for all counts of rank 1 and $\frac{1}{4}-\frac{1}{5} = \frac{1}{20}$th of the counts for the second bin. For each count of rank 2 we should, therefore, flip a `skewed' coin showing heads with probability $1/20$, and place it in the first bin if heads comes up, and in the second bin otherwise.

This procedure can be simplified as follows.
Consider ranks $r_1,...,r_n \in\{1,...,m+1\}$ and compute the transformed ranks
\begin{align}\label{transformation}
 \wt r_i := \frac{r_i - 1 + U_i}{m+1},
\end{align}
where $U_1,...,U_n$ are independent random variables, uniformly distributed on the interval [0,1].
The transformed ranks can take any value between 0 and 1, and we can now generate a histogram with any number of bins $k$ in the usual way, i.e. the $j$th bin counts the number of transformed ranks in the interval $[\frac{j-1}{k},\frac{j}{k})$. 
The random variables $U_i$ take the roles of the coinflips above, however, since they are uniformly distributed on [0,1] they automatically account for the fraction of overlap between the $k$ bins and the $m+1$ ranks.

The histogram of the modified ranks can be interpreted exactly as the original rank histogram. In fact, the randomization only has an effect if a bin number that does not divide $m+1$ is chosen, otherwise the two histograms are identical. 
After this replacement, histograms with any number of bins can be considered. Flatness is preserved and if the original ranks are uniformly distributed so are the transformed ranks.
Note that this also allows us to consider histograms with more than $m+1$ bins. If we, for example, consider $k = 2(m+1)$ bins, each count of rank 1 is simply assigned either to the first or to the second bin with equal probability.

This randomization is closely related to randomized versions of the probability integral transform (PIT), see e.g. \cite{Smith1985}.
When a probability forecast with distribution function $F$ is issued and observation $y$ materializes, the PIT simply considers $F(y)$. If the forecast system is reliable and $F$ is continuous, $F(y)$ follows a uniform distribution. Therefore a histogram of $F_1(y_1),...,F_n(y_n)$, for a sequence of observations and associated predictions, is a diagnostic tool for assessing the calibration of a probability forecast system, very similar to rank histograms for ensemble forecast systems. 
If the probability forecast $F$ is not continuous, \cite{Smith1985} suggested to modify the PIT by randomly filling in the jumps: That is, whenever the observation $y$ is at a discontinuity of $F$, the PIT value $F(y)$ is replaced by 
$F_-(y) + U(F_+(y) - F_-(y))$, where $F_-(y)$ and $F_+(y)$ are respectively the left
and right limits of $F$ at $y$.
 This modification allows in particular to consider the PIT for ensemble forecast systems by interpreting the ensemble forecast as its empirical distribution (resulting in a discontinuous distribution function with $m$ jumps). The resulting PIT histogram is then identical to the modified rank histogram suggested above.

As mentioned in the introduction, having with \eqref{transformation} a simple way of changing the number of bins in a histogram is useful in its own right. Especially when rank histograms are calculated on the same observations for competing forecast systems (with potentially different ensemble sizes), it is useful to make them comparable by creating histograms with the same bin number for both systems. Such a direct comparison can for example reveal if one of the two models is substantially more biased or underdispersed than the other. However, it is important to recognize that rank histograms are diagnostic tools and not designed for model comparison. As pointed out by \cite{Hamill2001}, flatness of histograms may result from mutual compensation between situations where the ensemble system is not reliable, and
observed flatness must be interpreted with caution.

\section{Tests for uniformity depending on the bin number}\label{sec:Theory}

In this section we review three tests for uniformity of the distribution of observation ranks, and consider the number of bins as an additional parameter in the test. This will allow us to adjust the bin number such that the test is approximated by a scientists' intuitive decision. It should be stressed that considering the bin number as a parameter is not useful from a data-analytic point of view: Reducing the number of bins by aggregating multiple observation ranks into the same bin constitutes a loss of information that generally reduces the power of a test for uniformity. Therefore, for assessing whether the observation ranks are uniformly distributed, statistical tests such as the $\chi^2$-test should be applied to the observation ranks directly, without aggregating them into fewer bins. 
Adjusting the number of bins used in a rank histogram is mostly relevant when histograms are used for intuitive inspection, i.e. as tools for visual diagnostics.

The three tests we consider are the classical $\chi^2$-test, a test based on the so-called reliability index \citep{DelleMonache&2006}, and a test considered by \cite{Taillardat&2016} based on an entropy statistic. We will refer to the latter two as RI-test and entropy test, respectively. For their formal definition, as well as a comparison of their performance, we refer to \cite{Wilks2019}. 
The tests are conceptually similar in that the test statistic is always a distance between the observed histogram and a perfectly flat histogram. The hypothesis of uniformity is rejected when this distance exceeds a threshold value, which is determined by the significance level of the test.
However, the tests differ in their definition of distance: the $\chi^2$-test is based on the $L^2$-distance, the RI-test is based on the $L^1$-distance, and the entropy test is based on the Kullback-Leibler-divergence.

In our context, it is convenient to rescale histograms such that their domain is the interval $[0,1]$ and integrate to a total area of one. In particular, we interpret rank histograms as histograms for data points distributed in the interval $[0,1]$, with the transformation \eqref{transformation} in mind. This simplifies notation greatly when considering different bin numbers for the same underlying data.
We generally denote the number of bins by $k$ and the height of the bins by $h_1,...,h_k$. Consequently, the frequency of the observation falling into the $j$th bin is $h_j/k$, and for a perfectly flat histogram we have $h_1 = \dots = h_k = 1$. For a histogram $H_k$ with $k$ bins we  then consider the three test statistics, or distances,
\begin{align*}
D_{L^2} := \frac 1 k \sum_{i = 1}^k (h_i-1)^2,\quad D_{L^1} := \frac 1 k \sum_{i = 1}^k |h_i-1|,\quad\text{and}\quad D_{KL} := \frac 1 k \sum_{i = 1}^k h_i \log(h_i),
\end{align*}
where for $D_{KL}$ we follow the convention that $0\log(0) = 0$.
The first two are the $L^2$- and $L^1$-distance between $H_k$ and a flat histogram, respectively. The third statistic is the Kullback-Leibler divergence from $P(H_k)$ to $U$, where $P(H_k)$ is the probability distribution defined by the bin frequencies of $H_k$, and $U$ is the uniform distribution. 

For each of these distances, a statistical test is obtained for the null hypothesis that the underlying data is uniformly distributed. That is, the null hypothesis is rejected if
\begin{align}\label{test}
D(H_k) > c(\alpha,k,n),
\end{align}
where $\alpha$ is the significance level of the test. The threshold $c(\alpha, k, n)$ is defined as the smallest value $c$ satisfying $P[D(H_k ) > c] \leq
\alpha $, when $H_k$ is a histogram (with $k$ bins) generated from $n$ independent uniformly distributed random variables. 
If we choose $D = D_{L^2}$, we recover the classical $\chi^2$-test, for $D = D_{L^1}$ we obtain the RI-test from \cite{DelleMonache&2006}, and, for $D = D_{KL}$, the entropy test from \cite{Taillardat&2016}.

We now aim to choose the bin number $k$ such that a scientist's intuitive decision approximates such a formal statistical test. To this end we make the following assumption, for all three distances, i.e. $D \in \{D_{L^2},D_{L^1},D_{KL}\}:$
\begin{enumerate}[(A)]
\item There is an `acceptance threshold' $\acct$ such that the scientist's intuitive decision is well-approximated by rejecting whenever $D(H_k)>\acct$. The acceptance threshold may depend on the chosen distance $D$.\label{enum1}
\end{enumerate}
Note that this assumption can be satisfied to different degrees for the different distances. It is, for example, possible that the use of an acceptance threshold constitutes a decent approximation to human behavior for $D = D_{L^2}$, but not for $D = D_{L^1}$.
To what extent this assumption is satisfied by the different distances is assessed in the next section, where we also use the results of an empirical study to derive reasonable values for $\acct$.

Subject to Assumption (A) being satisfied for one of the three distances $D_0$, we can choose the bin number such that the scientist's intuitive decision approximates the formal test based on $D_0$.
To this end, we choose a bin number $k$ such that $\acct \approx c(\alpha,k,n)$ from equation \eqref{test}. Then, by Assumption (\ref{enum1}), the scientist's decision is close to the statistical test. 
The derived bin number then depends on the number of available observations $n$ and on the significance level $\alpha$ of the test that is approximated. For a fixed number of observations $n$, the threshold $c(\alpha,k,n)$ is generally increasing in $k$ and decreasing in $\alpha$, see Section \ref{sec:Results}. Consequently, if $\alpha$ is chosen small, $k$ needs to be chosen small as well in order to achieve $\acct\approx c(\alpha,k,n)$. This is intuitive, since for a small significance level only a small probability of a false reject is allowed. Reducing the bin number generally leads to flatter histograms if the underlying data is uniformly distributed, and therefore reduces the chance of an intuitive false reject by the scientist.

 To sum up, in our proposed framework the optimal bin number $k_{opt}$ is the one that minimizes $|c(\alpha,k,n)-\acct|$. It depends on the number of available observations $n$, the selected significance level $\alpha$, and the acceptance threshold $\acct$. Such an optimal bin number can be derived for each of the three distances $D_{L^1},D_{L^2}$ and $D_{KL}$. Subject to Assumption (A), selecting this number of bins ensures that scientists' intuitive decisions are as close as possible to the statistical test associated with the corresponding distance.

\section{The acceptance threshold}\label{sec:c}

In this section we present the results of an empirical study assessing the validity of Assumption (\ref{enum1}) for the three different distances and derive approximations of the acceptance threshold. In this study several statisticians labeled histograms according to whether they believe them to be generated from uniform data or not. The histograms were in fact not based on underlying data at all, but were designed to have varying distances from uniformity. Further details of the study design are given in the appendix. More than 15 statisticians participated and 432 histograms were labeled.

\begin{figure}[h]
\includegraphics[width = 0.6\textwidth]{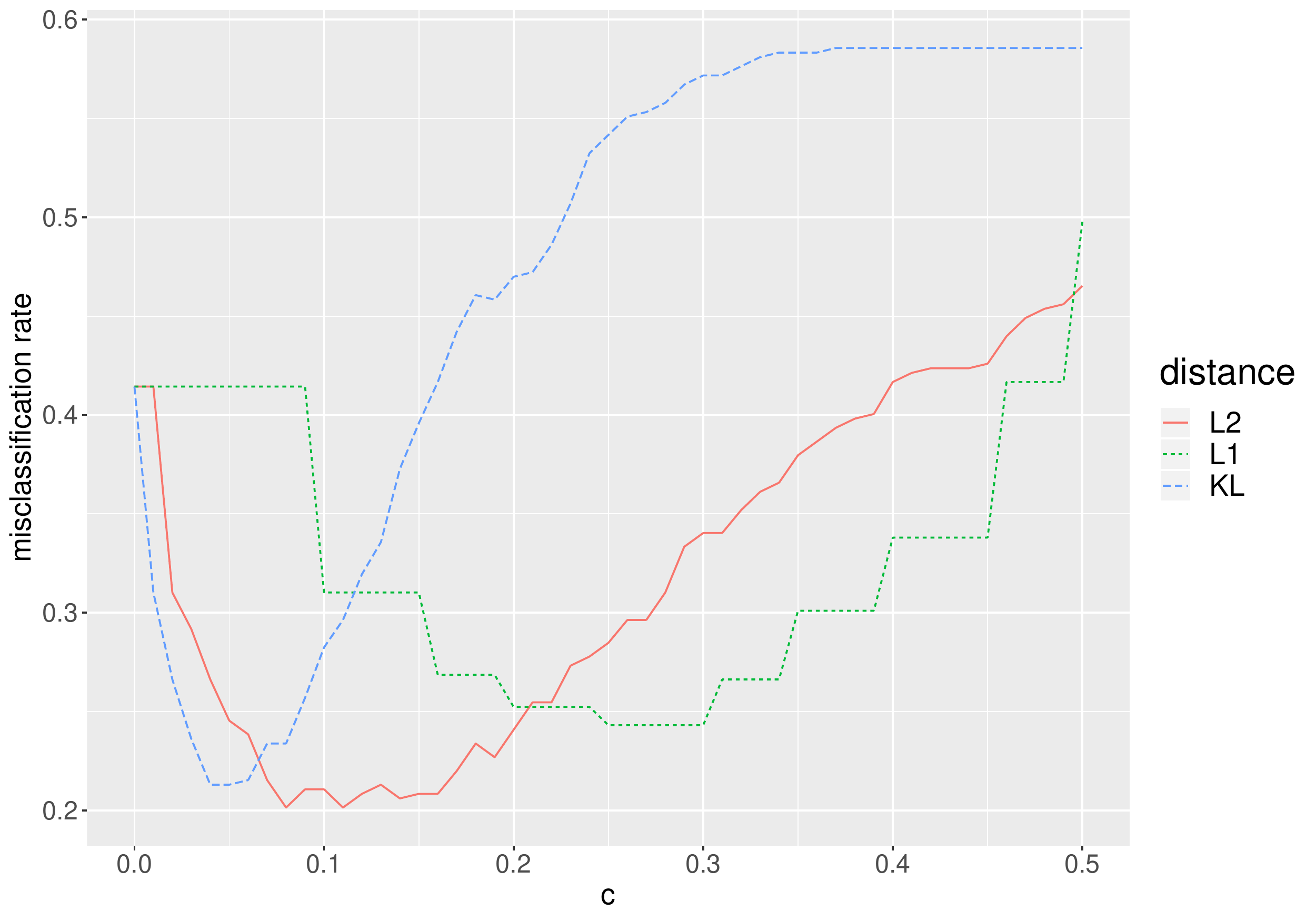}
\caption{The misclassification rate of the binary classifier $C_c$ for the three distances $D_{L^2}$, $D_{L^1}$ and $D_{KL}$, as a function of the acceptance threshold $c$. The low values all three curves attain at their minimum indicate that the classifier $C_c$ is a decent approximation for a scientist's intuitive decision, with $D_{L^2}$ and $D_{KL}$ providing slightly better approximations than $D_{L^1}$. 
The misclassification rate of $D_{L^1}$ is a step function due to the design of the empirical study: In a first version of this paper only the $L^1$-distance was considered, and the participants were therefore presented histograms that were generated to have a predefined $L^1$-distance, namely \{0.1,0.15,...\}. The distances $D_{L^2}$ and $D_{KL}$ of the labeled histograms were computed later on.
\label{figure:mcrate}}
\end{figure}

For $D\in\{D_{L^2},D_{L^1},D_{KL}\}$ we consider the binary classifier
\[C_c(D(H_k)) = 
\begin{cases}
\text{accept if $D(H_k)\leq c$,}\\
\text{reject if $D(H_k) > c$}
\end{cases}\]
and compare the decision of this classifier to the intuitive decisions made by the statisticians.
For a range of different $c$, we compute the misclassification rate of $C_c$, i.e. the proportion of cases where $C_c$ decided differently than the statistician. The value $c$ minimizing the misclassification rate then constitutes a good choice for $\acct$, and the misclassification rate at this value provides a measure for the extent to which Assumption (\ref{enum1}) is satisfied. 
The results for all three distances are shown in Figure \ref{figure:mcrate}. The lowest overall misclassification rate of 0.2 is achieved for $D = D_{L^2}$ and $c = 0.1$. In other words, rejecting a histogram whenever its $L^2$-distance exceeded 0.1 led to the same decision as the intuitive labeling for 4 out of 5 histograms. For $D_{KL}$ a similarly small misclassification rate was achieved, whereas the misclassification rate for $D_{L^1}$ was slightly higher, see Table \ref{table:cs} for details.

Different scientists have different preferences, and a histogram considered uniform by an optimist might be rejected by a pessimist. For the analysis in our next section we will therefore consider three different acceptance thresholds. The threshold minimizing the misclassification rate $\acct$, which provides the best fit to the results of our empirical study, as well as thresholds $c_{-}$ and $c_{+}$, representing a pessimist and an optimist, respectively. For all three distances, $c_-$ and $c_+$ were chosen such that the misclassification rate of $C_c$ with respect to our study results was approximately 5\% higher than for $\acct$.
The acceptance thresholds for the different distances and their corresponding misclassification rates are given in Table \ref{table:cs}.

\begin{table}
\centering
\begin{tabular}{l|cc|cc|cc}
			&	$\acct$	& {\it mcr}				&	$c_{-}$	& {\it mcr}			&	$c_{+}$	& {\it mcr}			\\	\hline
$D_{L^2}$	&	0.1		& 0.20					&	0.05	& 0.25				& 	0.2		&	0.24			\\
$D_{L^1}$ 	&	0.25	& 0.24					& 	0.15	& 0.31				&	0.35	&	0.30			\\
$D_{KL}$	&	0.05	& 0.21					& 	0.02	& 0.27				&	0.09	&	0.26
\end{tabular}
\caption{The three different values $\acct$, $c_{-}$ and $c_{+}$ considered as acceptance thresholds in Section \ref{sec:Results}, and their corresponding misclassification rates. The value $\acct$ is chosen to minimize the misclassification rate ($mcr$).  \label{table:cs}}
\end{table}

\begin{figure}[h]
\includegraphics[width = 0.95\textwidth]{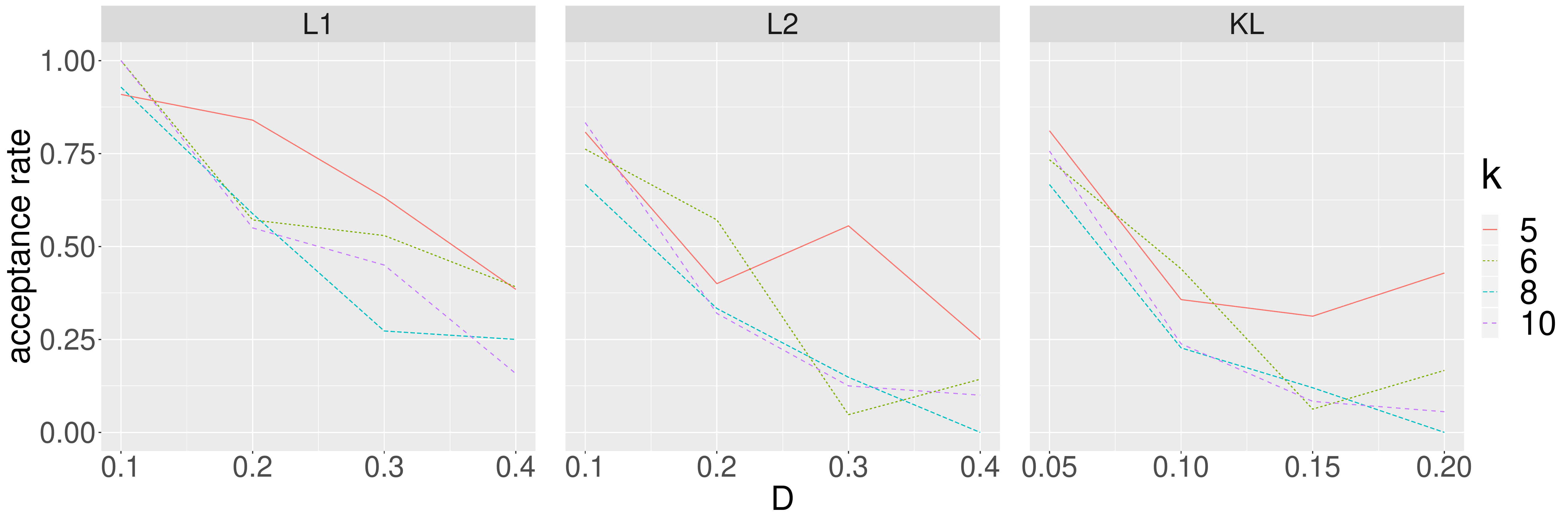}
\caption{The acceptance rate of the statisticians as a function of the distance, separately for different bin numbers $k$. For $D_{L^1}$ and $D_{L^2}$ the histograms are aggregated over intervals of length 0.1. As an example, the value shown at $D = 0.2$ is the acceptance rate over all histograms with a distance in the interval $(0.1,0.2]$. For $D_{KL}$ the same aggregation is applied over intervals of length 0.05.
\label{figure:acceptance}}
\end{figure}

In practice, the decision of an expert to accept or reject can depend on an interplay between a distance from uniformity and the number of bins $k$. For example, an $L^1$-distance of 0.25 for a histogram with 2 bins may be perceived as uniform, while the same distance of a histogram with 10 bins may be perceived as unacceptable.
 Such effects are unwanted in our context, since they are not accounted for by Assumption (\ref{enum1}). In order to control for this effect, the 432 histograms labeled in the study had different bin numbers, namely 5,6,8, or 10 bins.
Figure \ref{figure:acceptance} shows the acceptance rate of the scientists as a function of $D(H_k)$, for all three distances, and for each bin number $k$ separately. The figures suggest that, at the same distance from uniformity, histograms with fewer bins tend to have a slightly higher acceptance rate. This is also supported by the correlation between bin number and scientist's decision, which was -0.16 if acceptance by the scientist got assigned the value 1 and rejection got assigned the value 0.
This effect is particularly clear for large values of $D_{L^2}$ and $D_{KL}$ and for 5 bins. An explanation for this could be that both $D_{L^2}$ and $D_{KL}$ put a higher penalty on outlier-bins than $D_{L^1}$, which could indicate that the labeling scientists found outlier-bins more likely to occur when few bins were used. Overall, however, the effect of the bin number on the decision is small compared to the effect of the distance. 

\FloatBarrier
\section{Results}\label{sec:Results}

Here we present optimal bin numbers for a range of significance levels $\alpha$ and sample sizes $n$. 
As argued in the introduction, the results are mostly relevant for small data sizes $n$, and we restrict our analysis to $n\leq 200$. 
We compute the optimal bin number for all three distances and the acceptance thresholds $c_-,$ $\acct$ and $c_+$ given in Table \ref{table:cs}. For $\alpha$ we consider the classical choice of $5\%$, as well as the more relaxed choices $\alpha = 10\%$ and $\alpha = 33\%$. While in most scenarios a statistical test with a false rejection probability of $33\%$ is rather useless, such a threshold is not unreasonable in our informal setting where the test is approximated by scientists' intuitive decisions. 

For given values of $n,\alpha, c$ and any of the distances $D_{L^2},D_{L^1},D_{KL}$, the optimal number $k$ is then derived as follows. For all $k$ in the range $k=2,...,12$ we compute $c(\alpha,k,n)$ from \eqref{test} and choose $k$ such that $|c(\alpha,k,n) - c|$ is minimized. For the derivation of $c(\alpha,k,n)$ we do not rely on closed-form formulas (as in the original formulations of the tests), but use Monte-Carlo approximation with $N = 1.000.000$ samples. 
To be precise, we generate histograms $H_1,..., H_N$ with $k$ bins, each of which is based on $n$ independent uniformly distributed data points on $[0,1]$. For each histogram we compute $D(H_k)$ and obtain $c(\alpha,k,n)$ as 
the minimal value such that the fraction of histograms with $D(H_k)>c(\alpha,k,n)$ is smaller or equal to $\alpha$.

\begin{figure}[H]
\includegraphics[width = 0.9\textwidth, height = 0.8\textwidth]{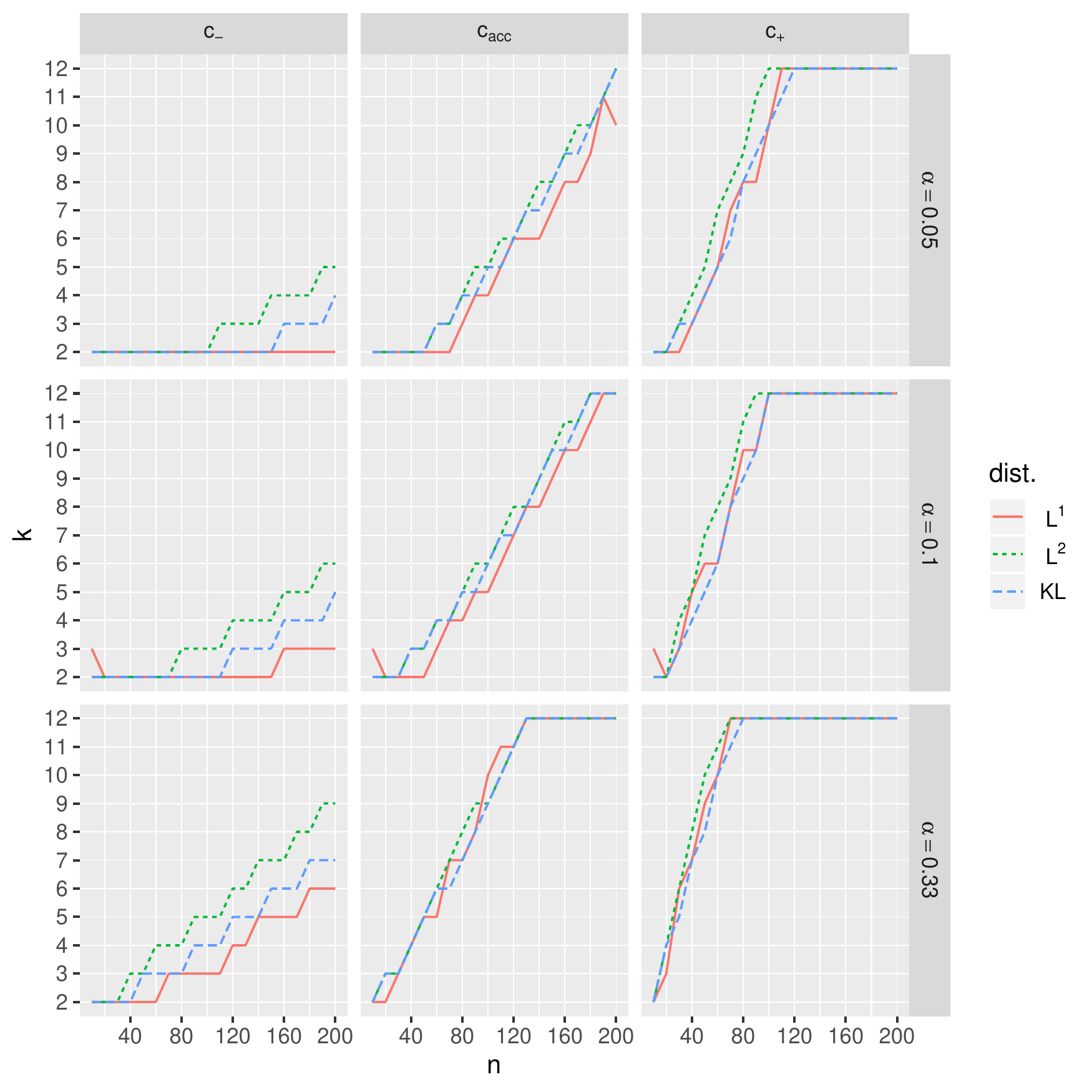}
\caption{The optimal bin number as a function of the data size $n$, for three different significance levels $\alpha$, and the three choices of acceptance threshold $c$, specified in Table \ref{table:cs}.
\label{figure:k_by_n}}
\end{figure}

The results are presented in Figure \ref{figure:k_by_n}. 
 It is clear to see that the bin number tends to increase in the sample size $n$ which is intuitive, since larger values of $n$ reduce the sample variability and therefore allow for separating the data into more bins. 
This effect is, nevertheless, remarkable since it is not obvious from the way the optimal bin number is derived.
Indeed, the occasional dips of the red curves in Figure \ref{figure:k_by_n} show that the increasing behavior in $n$ constitutes a tendency rather than a mathematical necessity. 
The increasing behavior can be explained by properties of the three distances used in the derivation. When the underlying data is uniformly distributed, the distance from uniformity of a histogram with fixed bin number $k$ tends to decrease when the number of data points $n$ increases. On the other hand, the distance from uniformity tends to increase if the number of bins $k$ is increased for a fixed sample size $n$. While this behavior is not directly shown in the figure, it implies that larger sample size $n$ is balanced by larger $k$, in order to keep the probability that the distance from uniformity exceeds the acceptance threshold at approximately $\alpha$, and therefore that the optimal bin number tends to increase in $n$.

The results differ strongly between the different acceptance thresholds $c_-$, $\acct$ and $c_+$, highlighting that the optimal bin number depends substantially on the preferences of the inspecting scientist. We will focus on the results for $\acct$, which provides the best approximation to our empirical study. Moreover, the study suggests that $D_{L^2}$ and $D_{KL}$ are better suited to approximate human behavior than $D_{L^1}$, which suggests to focus on the results for these two distances.
Furthermore, \cite{Wilks2019} concludes from his comparative analysis of the three tests that \emph{`the traditional $\chi^2$ test is recommended
as a consequence of its generally superior power, particularly for the underdispersed ensembles that are most}
\emph{commonly encountered, and the relative ease of obtaining the necessary critical values.'}
This suggests putting most emphasis on the bin numbers derived by using the $L^2$-distance.
There is remarkable similarity between the optimal bin numbers for $D_{L^2}$ and $D_{KL}$ when $c = \acct,$ which provides a sanity check for our approach: Even though the derivation of 
\begin{figure}[H]
\includegraphics[width = 0.9\textwidth, height = 0.8\textwidth]{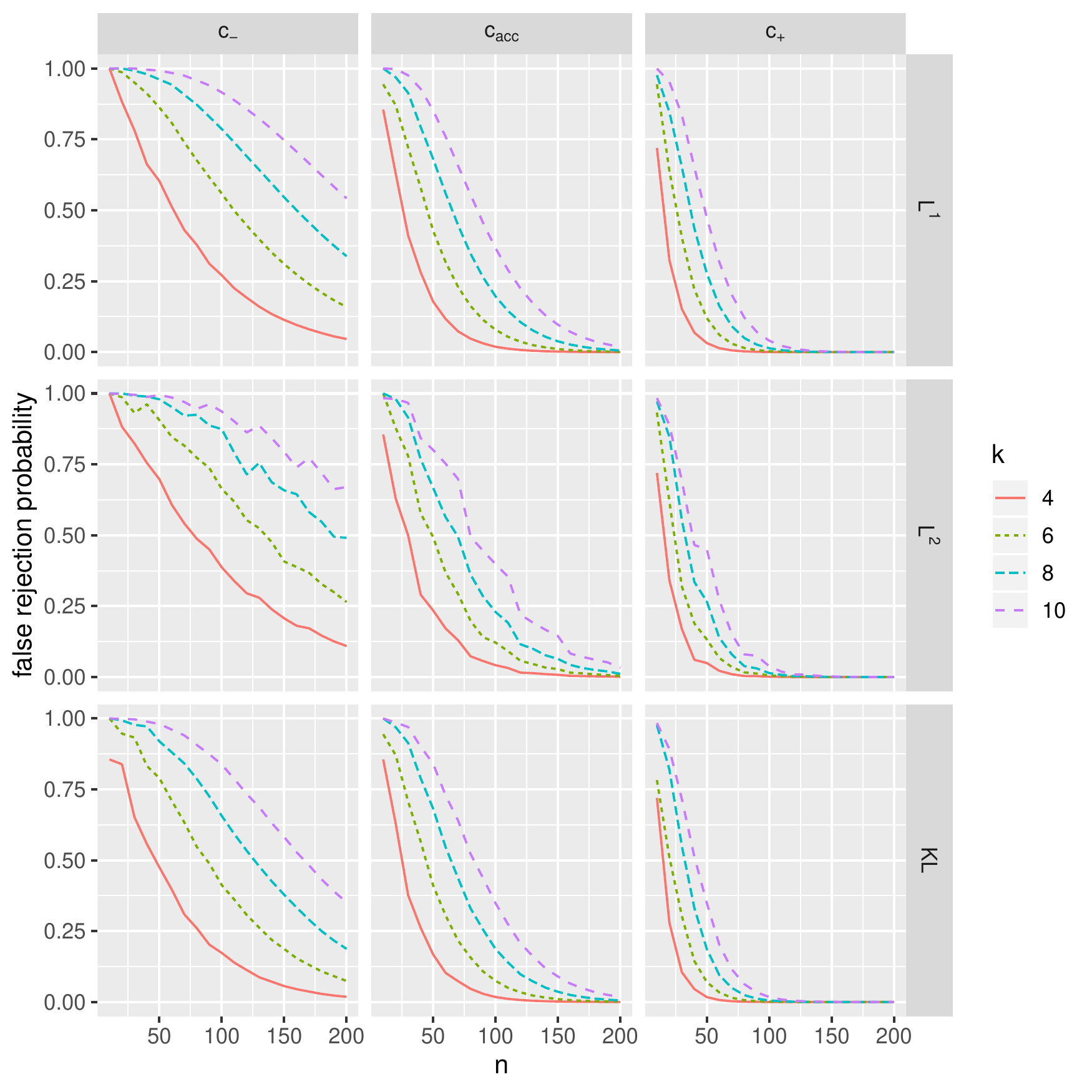}
\caption{The probability of a false rejection as a function of the data size $n$, for $k=4,6,8$ and $10$ bins.
Increasing the bin number leads to a higher probability for a false reject, but at the same time increases the probability for a correct reject if the underlying data is not uniformly distributed, cf. Figure \ref{fig:alternative_dists}.
\label{figure:frp_by_n}}
\end{figure}
the optimal bin number is based on different test statistics for different distances, the goal remains the same. Namely, to find a bin number that leads to an intuitive rejection of histograms of uniform data with probability $\alpha$.

As we would expect, the bin number $k$ increases not only in $n$ but also in $c$ and $\alpha$. The increase in $\alpha$ highlights that, if one is willing to accept large probabilities of a false reject, one should consider rank histograms with many bins, since this also tends to increase the probability of a correct reject (the power of the associated test) when the data is not uniformly distributed. The variability in $c$ mainly provides insight to what extent the results depend on the personal preferences of the scientist, but it should be mentioned that the selection of $c_-$ and $c_+$ in Section \ref{sec:c} is rather arbitrary.

Overall, the bin numbers suggested by this approach are relatively small, especially for small sample sizes $n$. For $n=100$, our approach suggests to choose only 5 bins in order to approximate a conservative test with significance level of 5\% (focusing on $\acct$ and either $D_{L^2}$ or $D_{KL}$). If we relax the significance level to 10\% (33\%), the algorithm selects 6 bins (9 bins) instead. In particular, if we have 100 forecast-observation pairs available, and we choose to print a histogram with 9 bins, we need to expect a roughly 33\% chance for an intuitive false reject if the ensemble forecast system is well-calibrated. If only 50 observations are available, the bin numbers drop to 2 (5\%), 3 (10\%) and 5 (33\%), respectively. Such bin numbers constitute a stark contrast to the common practice of choosing $m+1$ bins which typically results in 11 bins or more.
  
Instead of focusing on the theoretically optimal number of bins, we may analyze the false rejection rate of the classifier $C_c$ as a function of the bin number $k$. Figure \ref{figure:frp_by_n} shows the results for the bin numbers $k = 4,6,8$ and 10.
Again, we observe that the differences between the distances $D_{L^2}$, $D_{L^1}$ and $D_{KL}$ are small. Especially for the pessimistic threshold $c_-$ the false rejection probabilities are very large, even for small number of bins. This can be interpreted as a warning not to be too pessimistic when visually inspecting rank histograms based on few observations, but rather acknowledge that the natural variability is likely to result in histograms that may not look approximately flat, even when the underlying data is uniformly distributed.
\FloatBarrier

\section{Rejection probabilities under non-uniform distributions}\label{sec:non-uniform}

\begin{figure}[b]
\centering
\includegraphics[width = 0.49\textwidth]{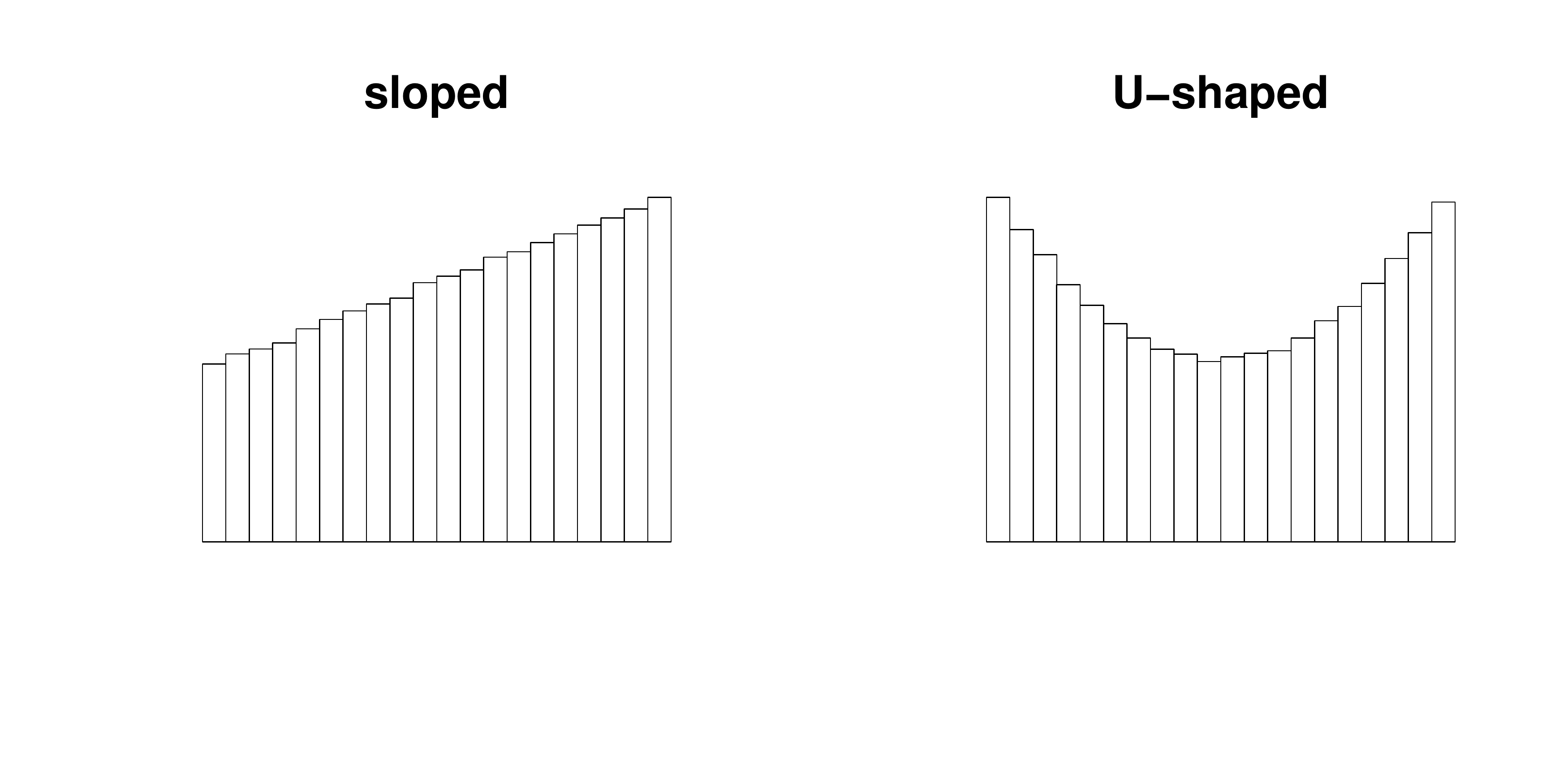}
\caption{Histograms of the two considered non-uniform distributions.\label{fig:alternative_dists}}
\end{figure}

In this section we analyze the rejection probability of the considered tests under non-uniform distributions. We consider two distributions representing the most prominent characteristic shapes that are important in rank histogram analysis.
The first distribution is sloped, with a density linearly increasing from 2/3 at 0 to 4/3 at 1, representing rank histograms based on a biased prediction system. The second distribution is U-shaped representing rank histograms based on an underdispersed prediction system. The U-shaped distribution has density $f(x) = 3(x-1/2)^2 + 3/4$, which is symmetric around 1/2 where it reaches its minimum value of 3/4.
Figure \ref{fig:alternative_dists} shows histograms of the two distributions based on 200.000 samples.

\begin{figure}
\centering
\includegraphics[width = 0.7\textwidth]{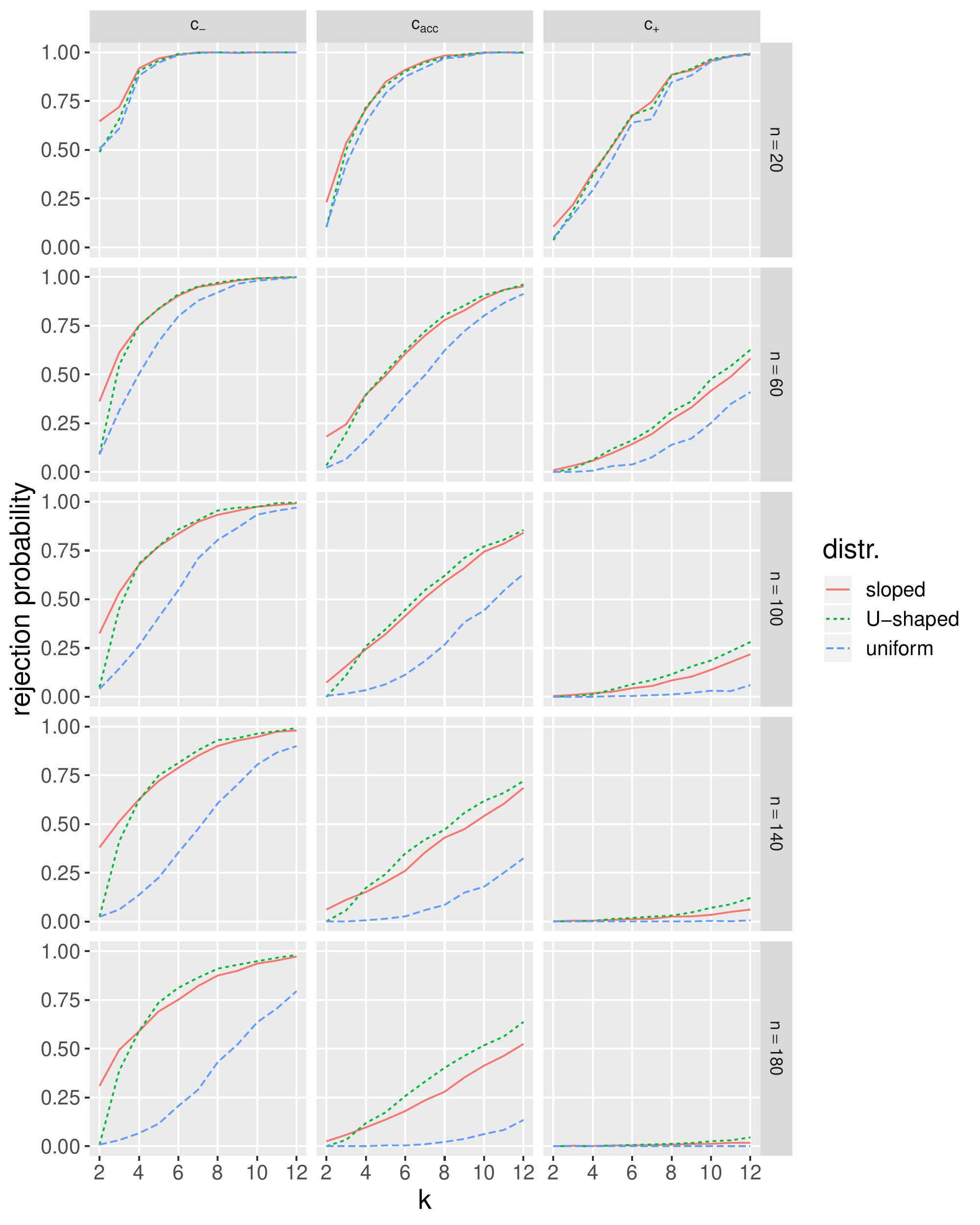}
\caption{Rejection probabilities for a range of $n$ and $k$ for the uniform distribution, and the sloped and the U-shaped distribution described in Section \ref{sec:non-uniform}. The results are only shown for the test based on the $L^2$-distance, and for the three acceptance thresholds $c_-,\acct,c_+$ given in Table 1.
\label{fig:power_by_k}}
\end{figure}

We obtain rejection probabilities for the three distributions by generating, for a range of $n$ and $k$, 1000 histograms with $k$ bins based on $n$ data points with the corresponding distribution,
and computing the distances $D_{L^1}, D_{L^2} $ and $D_{KL}$ for these histograms.
The rejection probability for one of these distances and a given acceptance threshold $c$ is then the fraction of histograms for which the distance exceeds $c$. 
As acceptance thresholds we consider the three values $c_-$, $\acct$ and $c_+$ specified in Table 1.
Figure \ref{fig:power_by_k} shows the rejection probabilities for these acceptance thresholds under the three distributions, for a range of bin numbers and sample sizes. The figure only shows the results for the $L^2$-distance, the other distances lead to very similar results (not shown). Generally, the rejection probability increases in the bin number, showing that histograms based on more bins tend to have a higher distance from uniformity under all three considered distributions.
The uniform distribution gets rejected with the lowest probability, which indicates that the considered tests are unbiased. However, when $k=2$, the U-shaped histogram gets rejected with the same probability. This highlights that histograms based on two bins are essentially useless in practice, since they cannot indicate misspecified dispersion in the ensemble forecast system.

The figure clearly visualizes the trade-off that is made in choosing the number of bins: While a low rejection probability is desirable when the data is uniformly distributed, high rejection probabilities are desirable for the two alternative distributions.
Figure \ref{fig:power_by_k} shows that using $c_+$ generally leads to very low rejection probabilities, even for non-uniform data.
The pessimistic threshold $c_-$, on the other hand, generally leads to much lower rejection probabilities for uniformly distributed data than for data generated from the alternative distributions. However, the probability for a false reject is generally very large when $c_-$ is used, for example it is more than 75\% when 12 bins are chosen, even for $n = 180$. The threshold $\acct$ suggested by our empirical study leads to a large difference in acceptance probabilities between uniform and non-uniform distributions and, at the same time, allows for reasonably small false rejection probabilities. 
It can generally be observed that the differences in rejection probabilities between uniform and non-uniform distribution are getting more clearly pronounced as $n$ increases. This highlights the fact that with more available data it becomes easier to differentiate between uniform and non-uniform distributions. It is also worth mentioning that the optimist's acceptance threshold $c_+$ performs reasonable well for $n<100$. Consequently, for very small $n$, one should be careful not to expect too uniform histograms.

 \begin{figure}
\centering
\includegraphics[width = 0.9\textwidth]{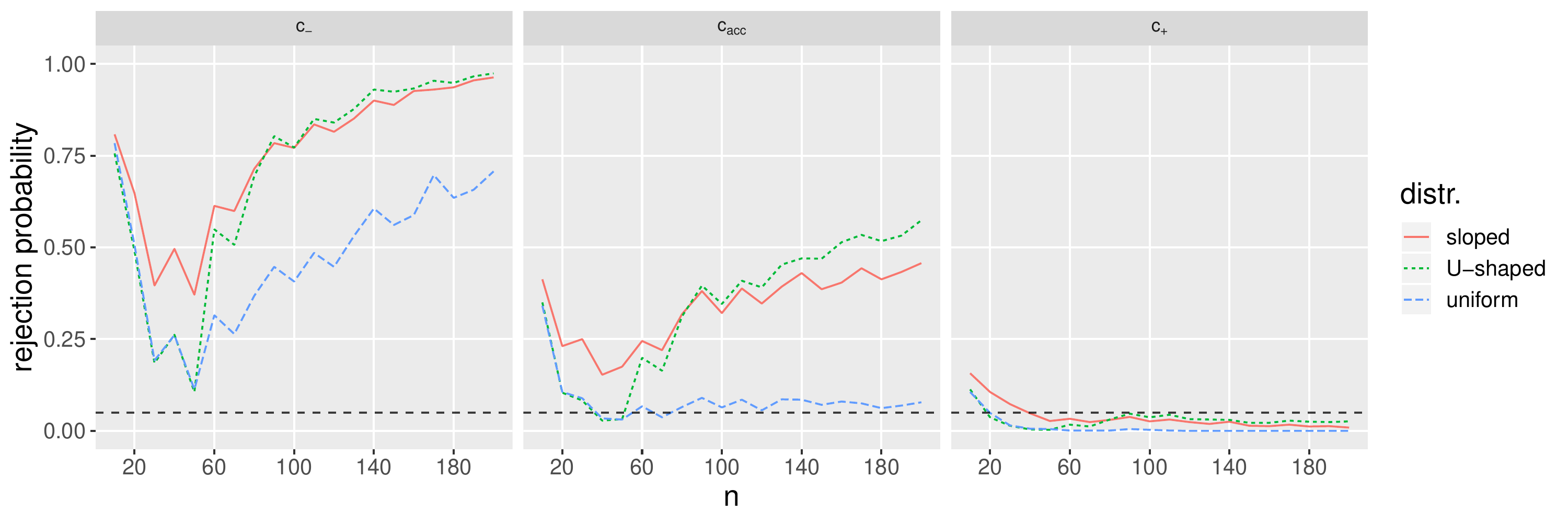}
\caption{Rejection probabilities for a range of $n$ when the optimal bin number is used.
\label{fig:power_opt_k}}
\end{figure} 
Figure \ref{fig:power_opt_k} shows the rejection probability for the three distributions when the optimal bin number is used. Here, the optimal bin number is derived using the $L^2$-distance, the acceptance threshold $c_{acc}$ and the significance level $\alpha = 5\%$. The significance level is shown in the figure as dashed line. The plot in the middle shows that the bin number is selected in order to align the blue line with the 5\% significance level. Note that approximately $n=40$ is required in order to achieve a false rejection rate of only 5\%, even when only two bins are used. 
The left hand side and right hand side plot show the rejection probabilities for pessimist and optimist, respectively, when they inspect histograms based on the optimal number of bins derived with the acceptance threshold $\acct.$

\section{Discussion}\label{sec:Discussion}

Our study indicates that, when visually inspecting forecast calibration with rank histograms, choosing a small number of bins can substantially lower the risk of wrongfully rejecting the hypothesis that the underlying data is uniform.

In practice, rank histograms are applied to identify characteristic shapes indicating certain miscalibrations of the ensemble forecast. This has several implications.
The most common characteristic shapes in the appearance of rank histograms are slopes (indicating bias) as well as $\cup$- and $\cap$-shapes (indicating under- and overdispersion, respectively). In particular, it is never advisable to only use two bins (as our approach suggests in some cases for very small sample sizes), since such a histogram is unable to pick up on dispersion misspecification.
At the same time, these simple shapes are equally well captured by a histogram with three bins than by histograms with many bins. More involved characteristic shapes (e.g. S-shapes) can indicate misspecified skewness or combinations of bias and misspecified dispersion. However, they often require a large sample size $n$ to become clearly visible, see \citet{ThorarinsdottirSchuhen2018}.
Such shapes are generally captured by histograms with six or eight bins, and it is difficult to imagine any informative characteristic shape that would require more than 10 bins in order to become visible. 
On the contrary, our results indicate that increasing the bin number puts more emphasis on random fluctuations in the data which can distract from characteristic shapes. Based on these considerations we recommend to generally
limit the number of bins in histograms to about 10. When the number of available forecast-observation pairs is limited one should not hesitate to consider histograms with fewer bins. Histograms with three bins might look somewhat unusual, but may be more appropriate when $n$ is very small in order to mitigate effects of sampling uncertainty.

At the same time, choosing a very small number of bins increases the risk of not recognizing deviations from uniformity, as shown in Section \ref{sec:non-uniform}. Moreover, in situations where the size of the verification data set is not known to the inspector, a larger number of bins can help the inspector to estimate how many forecast-observation-pairs were used and thus to avoid false acceptance or rejection of uniformity.

We assumed throughout this paper that the ranks of the different forecast-observation-pairs are independent. This assumption is commonly made when rank histograms are constructed, but
is violated in some applications, in particular when multiple spatial grid points are considered as samples. 
Such complex dependence structure can make the histogram much harder to interpret and, in particular, prevent formal testing for uniformity. See \citet{Hamill2001} for an in-depth discussion of this topic.

\section{Conclusion}\label{sec:Conclusion}

We introduce a criterion for choosing the number of bins in a rank histogram. The criterion attempts to make the intuitive decision of scientists regarding calibration close to a statistical test. 
It addresses the trade-off that adding more bins leads to a more detailed histogram but at the same time 
decreases statistical robustness, and attempts to optimize intuitive decision making based on the histogram. Our results highlight that the probability for intuitively rejecting a histogram tends to increase with the number of bins, even if the underlying data is uniformly distributed. This generally questions the current practice of choosing as many bins as possible. We showed that reducing the bin number can, to some extent, be used to appropriately balance the probability of an intuitive false reject, which also depends on the sample size $n$. This probability further depends on the preferences and experience level of the inspecting scientist. The bin numbers derived in the previous section are therefore merely suggestions based on our empirical study and do not constitute theoretical optima that ought to be followed under all circumstances.

Our results indicate that, especially for small verification samples with less than 100 data points, histograms with five bins or fewer are preferable. If histograms with more bins are considered, their appearance should not be over-interpreted, and rather large deviations from flatness should be expected, even for histograms based on uniformly distributed data. 
Moreover, for very small sample sizes of 50 or less, the probability for an intuitive false reject is generally rather large (often $50\%$ or higher), for any reasonable bin number ($k>2$). This highlights the large uncertainty associated with such small sample sizes and shows that rank histograms should in such situations be interpreted very carefully.
Generally, and particularly in this case, rank histogram analysis should rely on the results of statistical tests for uniformity rather than on intuitive inspection of the histogram plot. 
The importance of this is highlighted by our study that showed that intuitive decisions are strongly dependent on the selected number of bins, which is a property of the histogram plot only, not of the distribution of observation ranks in the predictive ensemble.

This article is accompanied by the \texttt{R}-package \Rpack which is available on the authors github account \github. The package includes functionality to generate histograms with any bin number from observed ranks using the transformation \eqref{transformation}, and to compute the optimal bin number for any sample size $n$, acceptance threshold $c$ and test size $1-\alpha$. 
Moreover, it provides tools and guidance that allow the reader to conduct the empirical study described in Section \ref{sec:c}. By personally labeling histograms you can derive your personal acceptance threshold $\acct$, and derive optimal bin numbers for histograms inspected by yourself.

\section*{Appendix: Details on the Empirical Study}

\begin{figure}[h]
 \centering
\includegraphics[width = 0.5\textwidth, height = 0.5\textwidth]{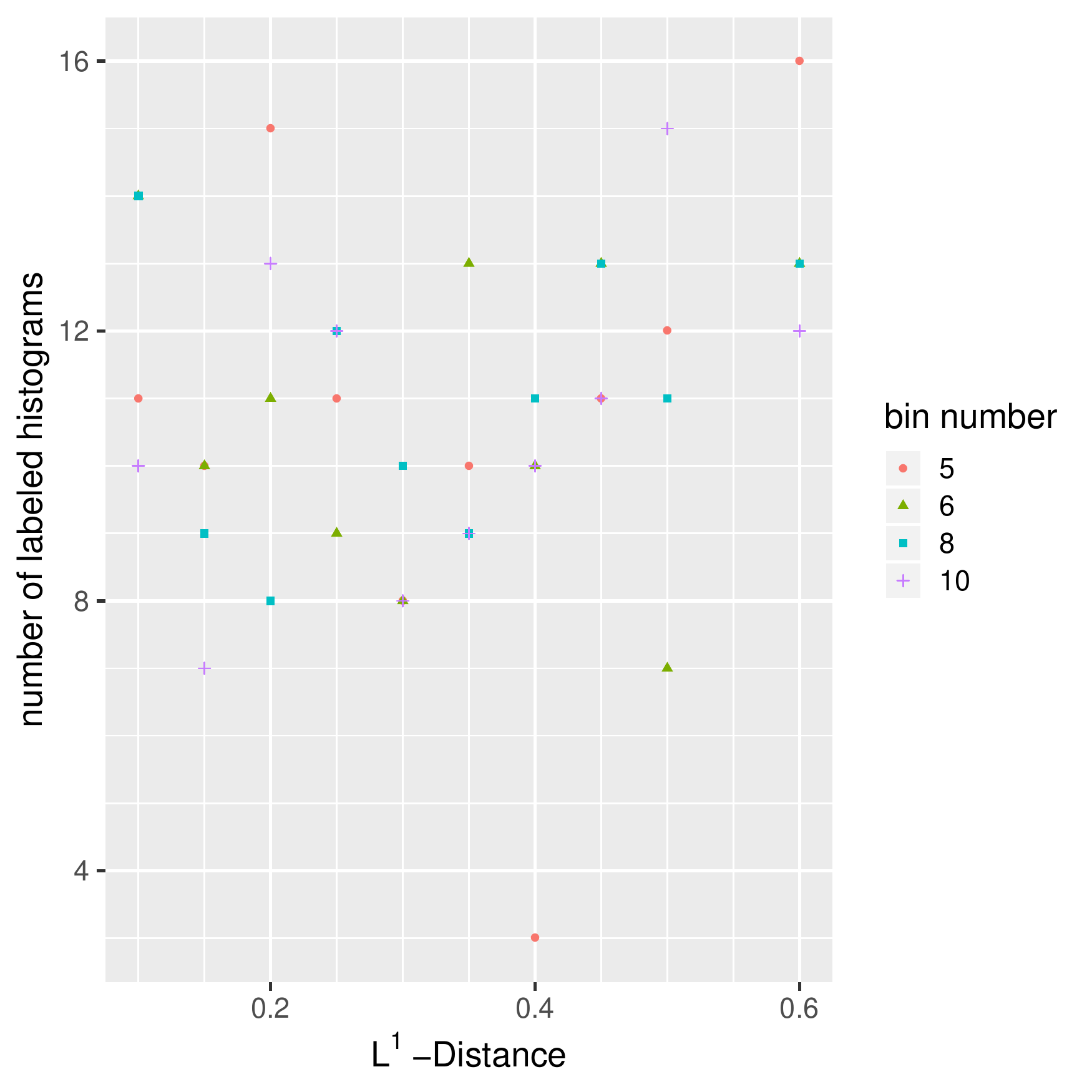}
\caption{How many histograms (out of 25 possible) were labeled in each category in the empirical study.
\label{fig:Nlabels}}
\end{figure}

Here we give more details about the design of the empirical study presented in Section 4. 
An early version of this paper only considered the $L^1$-distance from uniformity. Therefore, the study originally focused on analyzing the effect of different $L^1$-distances only. The analysis of $L^2$-distance and Kullback-Leibler divergence was added later and not taken into account for study design. For the study, 1000 histograms were created with 5,6,8 or 10 bins, and with $L^1$-distance in $\{0.1,0.15,...,0.45,0.5,0.6\}$. The histograms were not based on underlying data, but were sampled by an algorithm described below that allows to generate histograms with pre-specified number of bins and $L^1$-distance. Considering 4 different bin numbers and 10 different $L^1$-distances resulted in 40 categories, for each of which 25 histograms were created. The created histograms were shuffled, printed out and laid out in the break room of the statistics and data science group of the Norwegian Computing Center in Oslo, Norway, with a call to the group to label as many histograms as possible. The participants labeled the histograms according to whether they believe them to be based on uniform data or not, and were left unaware that the histograms were not based on underlying data at all.
The labeling of histograms was anonymous and participants could label as many histograms as they wanted. More than 15 Statisticians confirmed that they participated, and 432 out of the 1000 printed histograms were labeled. In all 40 categories the number of labeled histograms was between 7 and 16 (out of 25), except for one category where only three histograms were labeled. A detailed key of how many histograms were labeled in which category is shown in Figure \ref{fig:Nlabels}. 
 
The following algorithm was used for creating a random histogram with pre-specified bin number $k$ and $L^1$-distance from uniformity $D$.
\begin{enumerate}
\item Choose a number of steps $n$ (in the study $n = 50$) for the algorithm. Start out with a perfectly uniform histogram with $k$ bins. Mark all the bins with a 0.
\item Randomly select one of the bins marked 0 or 1, and \emph{increase} its height by $\frac{Dk}{2n}$. If the bin was marked 0, change its mark to 1.
\item Randomly select one of the bins marked 0 or -1, and \emph{decrease} its height by $\frac{Dk}{2n}$. If the bin was marked 0, change its mark to -1.
\item Repeat steps 2 and 3 in total $n$ times.
 \end{enumerate}
 In this algorithm, both steps 2 and 3 increase the $L^1$-distance from uniformity by $D/2n$, and, since they are both repeated $n$ times, the final histogram has $L^1$-distance from uniformity $D$. The marking is important to ensure that bins that have been increased (decreased) in height will only ever be increased (decreased), which ensures that the distance in fact increases in each step. The alternation between increasing and decreasing bin heights ensures that the total integral of the histogram remains 1.
 The algorithm needs an additionally constraint that prevents that bin heights are decreased beyond zero.



\end{document}